\documentclass[epj, twocolumn]{webofc}
\usepackage[varg]{txfonts}   
\usepackage{subfigure}
\usepackage{xspace}

\def\EPOS{\textsc{Epos}\xspace}
\def\EPOSfull{\textsc{Epos1.99}\xspace}
\def\SIBYLL{\textsc{Sibyll}\xspace}
\def\SIBYLLfull{\textsc{Sibyll2.1}\xspace}
\def\QGSJETIIfull{\textsc{QGSJetII-03}\xspace}

\newcommand{\eV}{\ensuremath{\mbox{e\kern-0.08em V}}\xspace}
\newcommand{\TeV}{\ensuremath{\mbox{Te\kern-0.08em V}}\xspace}
\newcommand{\GeV}{\ensuremath{\mbox{Ge\kern-0.08em V}}\xspace}
\newcommand{\MeV}{\ensuremath{\mbox{Me\kern-0.08em V}}\xspace}
\newcommand{\GeVc}{\ensuremath{\mbox{Ge\kern-0.08em V}\!/c}\xspace}
\newcommand{\MeVc}{\ensuremath{\mbox{Me\kern-0.08em V}/c}\xspace}

\newcommand{\Fluka}{{\scshape Fluka}\xspace}
\newcommand{\UrqmdLong}{{\scshape U}r{\scshape qmd1.3.1}\xspace}
\newcommand{\Urqmd}{{\scshape U}r{\scshape qmd}\xspace}

\newcommand{\Gheisha}{{\scshape Gheisha}\xspace}
\newcommand{\Corsika}{{\scshape Corsika}\xspace}
\newcommand{\Venus}{{\scshape Venus}\xspace}

\woctitle{International Symposium on Very High Energy Cosmic Ray
  Interactions (ISVHECRI 2012)}
\begin{document}
\selectlanguage{english}

\title{Results from NA61/SHINE}

\author{M.~Unger\inst{1}\fnsep\thanks{\email{Michael.Unger@kit.edu}}
for the NA61/SHINE Collaboration\fnsep\thanks{{https://na61.web.cern.ch/na61/pages/storage/authors\_list.tex}}}

\institute{Institut f\"ur Kernphysik, Karlsruher Institut für
  Technologie, Postfach 3640, D - 76021 Karlsruhe }

\abstract{
 In this paper we summarize recent results from NA61/SHINE relevant
 for heavy ion physics, neutrino oscillations and the interpretation
 of air showers induced by ultra-high energy cosmic rays.}
\maketitle
\section{The NA61/SHINE Experiment}
\label{intro}
NA61/SHINE (SHINE = SPS Heavy Ion and Neutrino Experiment)~\cite{na61}
is a multi-purpose fixed target experiment to study hadron production
in hadron-nucleus and nucleus-nucleus collisions at the CERN Super
Proton Synchrotron (SPS). Among its physics goals are precise hadron
production measurements for improving calculations of the neutrino
beam flux in the T2K neutrino oscillation experiment~\cite{Abe:2011ks}
as well as for more reliable simulations of hadronic interactions in
air showers. Moreover, $p$+$p$, $p$+Pb and nucleus+nucleus collisions
are measured to allow for a study of the properties of the onset of
de-confinement and a search for the critical point of strongly
interacting matter (see e.g.\ Ref.~\cite{Gazdzicki:2010iv}).

\begin{figure}[h]
\centering \includegraphics[clip, viewport= 50 0 475 430,
  width=\linewidth]{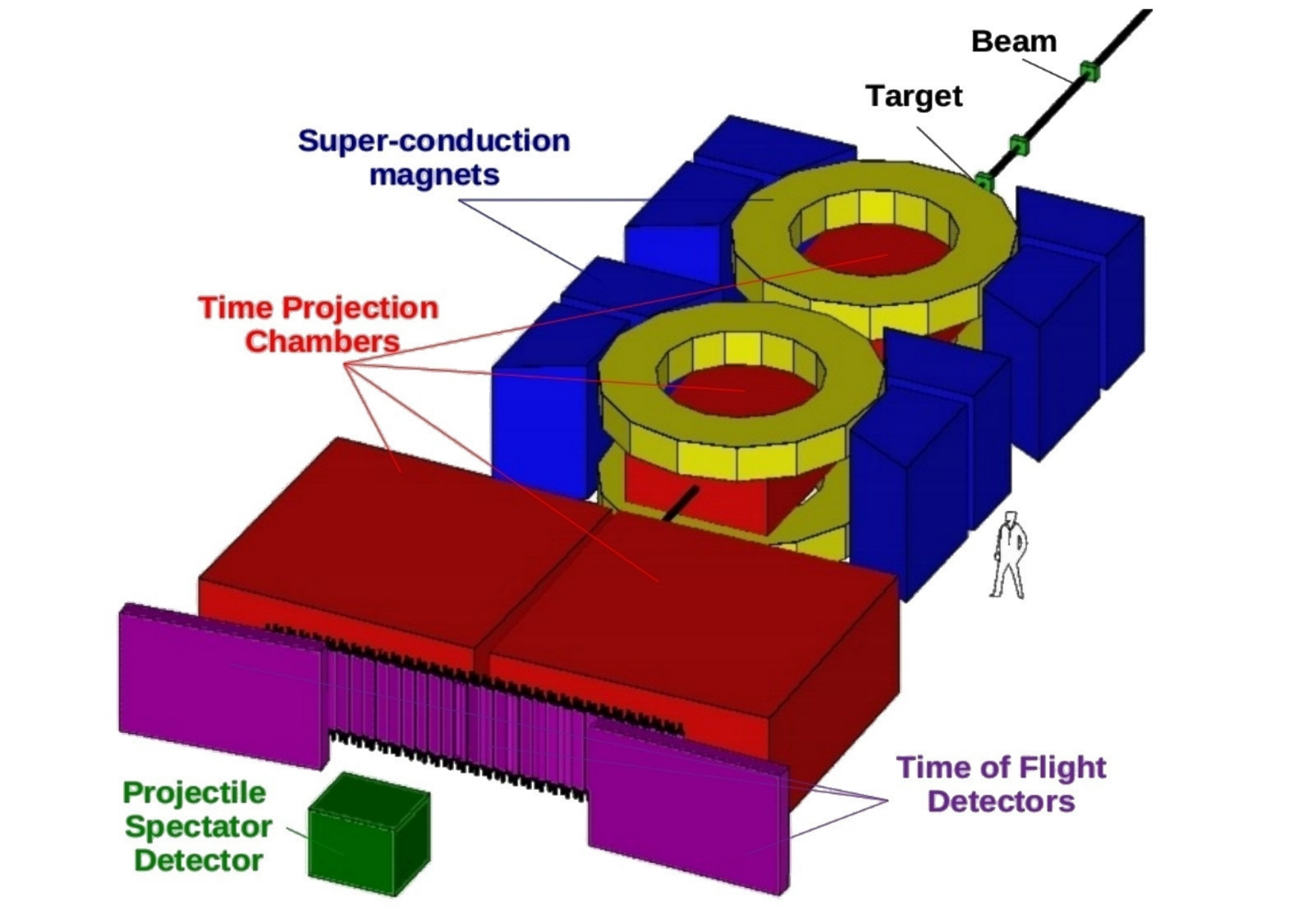}
\caption{Schematic layout of the NA61/SHINE experiment.}
\label{fig:setup}
\end{figure}

The layout of the NA61/SHINE detector is sketched in
Fig.~\ref{fig:setup}.  A set of scintillation and Cherenkov counters
as well as beam position detectors upstream of the spectrometer
provide timing reference, identification and position measurements of
the incoming beam particles.  Large time-projection-chambers (TPCs)
inherited from the NA49 experiment~\cite{Afanasev:1999iu} are used to
measure the charge and momentum of particles. The momentum resolution,
$\sigma(1/p)=\sigma(p)/p^2$, is about $10^{-4}$~(\GeVc)$^{-1}$ at full
magnetic field and the tracking efficiency is better than 95\%.
Particle identification is achieved by measuring the energy loss along
the tracks in the TPCs and by determining their velocity from the time
of flight provided by large scintillator walls placed downstream of
the TPCs.  The centrality of nucleus-nucleus collisions can be
estimated using the measurement of the energy of projectile spectators
with a calorimeter~\cite{Ivashkin:2012fd} located behind the time of
flight detectors. For nucleon-nucleus collisions, the centrality is
determined by counting low momentum particles from the target (so
called 'gray protons') with a small TPC around the target.

NA61/SHINE started data taking in 2007.  After a first run with proton
on carbon at 31~\GeVc, the data acquisition system was upgraded during
2008 to increase the event recording rate by a factor of $\approx
10$. In the last four years, a wealth of data has been recorded by the
experiment at beam momenta ranging from 13 to 350~\GeVc with various
beam particles and targets. In this paper we present results obtained
from the data relevant for heavy ion physics, neutrino oscillations
and the interpretation of air showers at ultra-high energies.
\section{Measurements of $p$+C Interactions for the Improvement
of Neutrino Flux Calculations}
Measurements of the particle emission from targets used to create
neutrino beams are important for a precise interpretation of
long-baseline neutrino oscillation experiments such as
Tokai-to-Kamioka (T2K)~\cite{Abe:2011ks}. Two types of measurements
have been performed by NA61/SHINE to aid the T2K calculations of the
neutrino fluxes: interactions of proton on a replica of the T2K target
(a 90~cm long graphite rod) and thin (2~cm) target measurements to
allow the measurement of single proton-carbon interactions. Both
measurements were performed with a 31~\GeVc proton beam, similar to
the one provided at J-PARC.  A total of 0.2$\times$10$^6$ events were
recorded during the data taking in 2007 and more statistics were
collected in 2009 (4$\times$10$^6$ events) and 2010 (10$\times$10$^6$
events). For the first time, the kinematic phase space of pions and
\begin{figure}[t]
\centering \includegraphics[clip, viewport= 5 0 530 430,
  width=\linewidth]{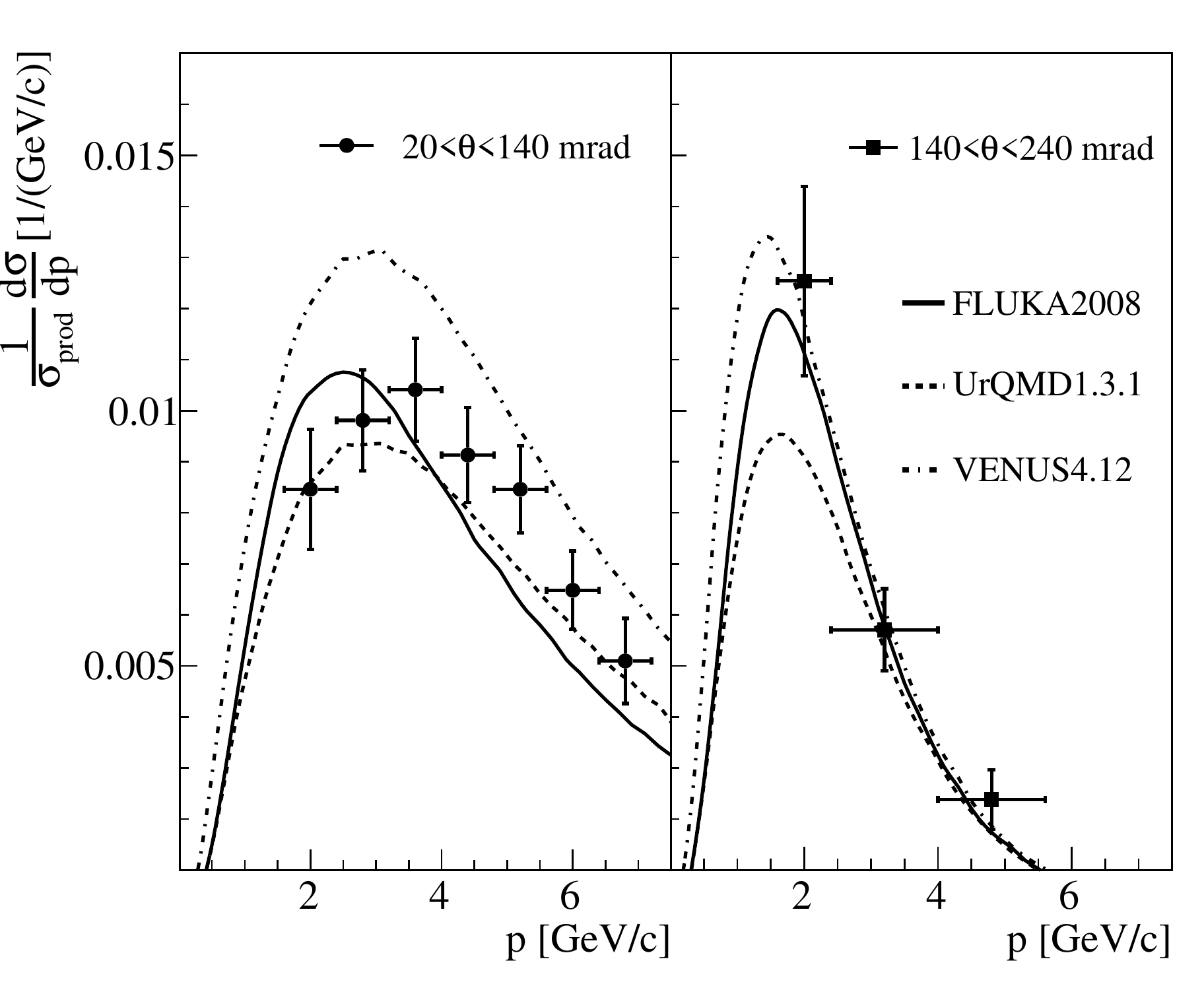}
\caption[kaon]{Comparison of measured $K^+$ spectra in $p$+C
  interactions at 31~\GeVc with model predictions.  The vertical error
  bars on the data points show the total (stat.\ and syst.)
  uncertainty. The lines indicate predictions using
  \Venus~\cite{Venus}, \Fluka~\cite{Fluka} and \Urqmd~\cite{Urqmd,
    Uzhinsky:2011ir}.}
\label{fig:kaon}
\end{figure}
kaons exiting the target and producing neutrinos in the direction of
the near and far detectors of a long-baseline neutrino oscillation
experiment is fully covered by a single hadron production experiment.

First results on pion and kaon yields have already been
published~\cite{NA61pion, Abgrall:2011ts} and used in the T2K data
analysis~\cite{Abe:2011sj}. An example of a recent particle yield
measurement for T2K is the inclusive yield of positive kaons in p+C
interactions at 31~\GeVc which is shown Fig.~\ref{fig:kaon}. The
knowledge of charged kaon yields is important for T2K because kaons
generate the high energy tail of the neutrino beam and contribute
substantially to the intrinsic $\nu_{e}$ component. As can be seen,
none of the superimposed model predictions can fully describe the
small-angle data from 20 to 140 mrad. Because of these shortcomings of
hadronic interaction models and similar deficits in case of a
comparison of predicted pion yields to NA61/SHINE data
(cf.~\cite{NA61pion}), neutrino flux predictions cannot be used
\begin{figure}[t]
\centering \includegraphics[clip, viewport=0 -40 520 390,
  width=\linewidth]{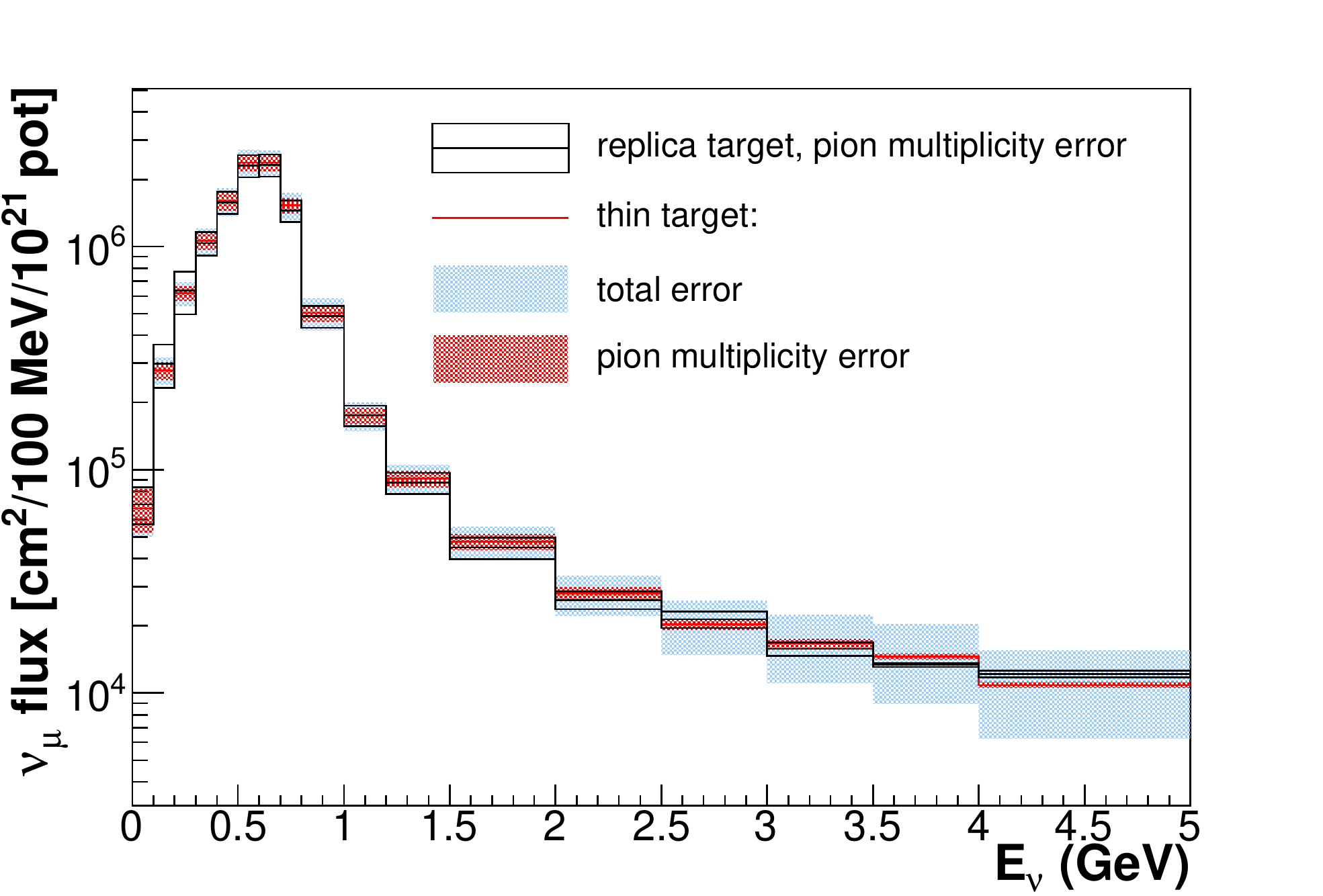}
\caption[flux]{Re-weighted $\nu_{\mu}$ flux predictions at the far
  detector of T2K based on the NA61/SHINE thin-target and replica-target
  data.}
\label{fig:flux}
\end{figure}
directly to interpret oscillation data, but need to be modified to
match the NA61/SHINE measurements. For this purpose flux simulations
are re-weighted to match the measured secondary particle yields,
either on an interaction-by-interaction basis using the thin-target
data, or at the surface of the T2K target by using data collected with
the replica-target~\cite{Abgrall:2012pp}. As can be seen in
Fig.~\ref{fig:flux}, the resulting calculated neutrino spectra at the
T2K far detector are in excellent agreement using either of these two
methods.

\section{NA61/SHINE Results for the Interpretation of Cosmic Ray Air Showers}
Cosmic rays initiate extensive air showers (EAS) when they collide
with the nuclei of the atmosphere.  The interpretation of EAS data as
for instance recorded by the Pierre Auger Observatory~\cite{Auger},
KASCADE~\cite{KASCADE} or IceTop~\cite{IceTop} relies to a large
extent on the understanding of these air showers and specifically on
the correct modeling of hadron+air interactions that occur during the
shower development. The relevant particle energies span a wide range
from primary energies of $\gtrsim 10^{20}$~\eV down to energies
of~$10^{9}$~\eV. The mesons that decay to muons at ground level
typically originate from low energy interactions in the late stages of
an air shower. Depending on the primary energy and detection distance,
the corresponding interaction energies are between 10 and 1000~\GeV
and the modeling of the corresponding low energy interactions
contribute at least 10\% to the overall uncertainty of the predicted
muon number at ground (see e.g.\ Refs.\ \cite{Heck:2003br,
  Drescher:2003gh, Meurer:2005dt, Maris:2009uc}).

Unfortunately, there exist no comprehensive and precise particle
production measurements for the most numerous projectile in air
showers, the $\pi$-meson. Therefore, new data with pion beams at 158
and 350~\GeVc on a thin carbon target (as a proxy for nitrogen) were
collected by the NA61/SHINE experiment at the CERN SPS and preliminary
results from this data set were presented at this conference for the
first time.
\begin{figure*}[t!]
\centering \subfigure[$h^-$ at 158\,\GeVc]{
  \includegraphics[width=0.485\linewidth]{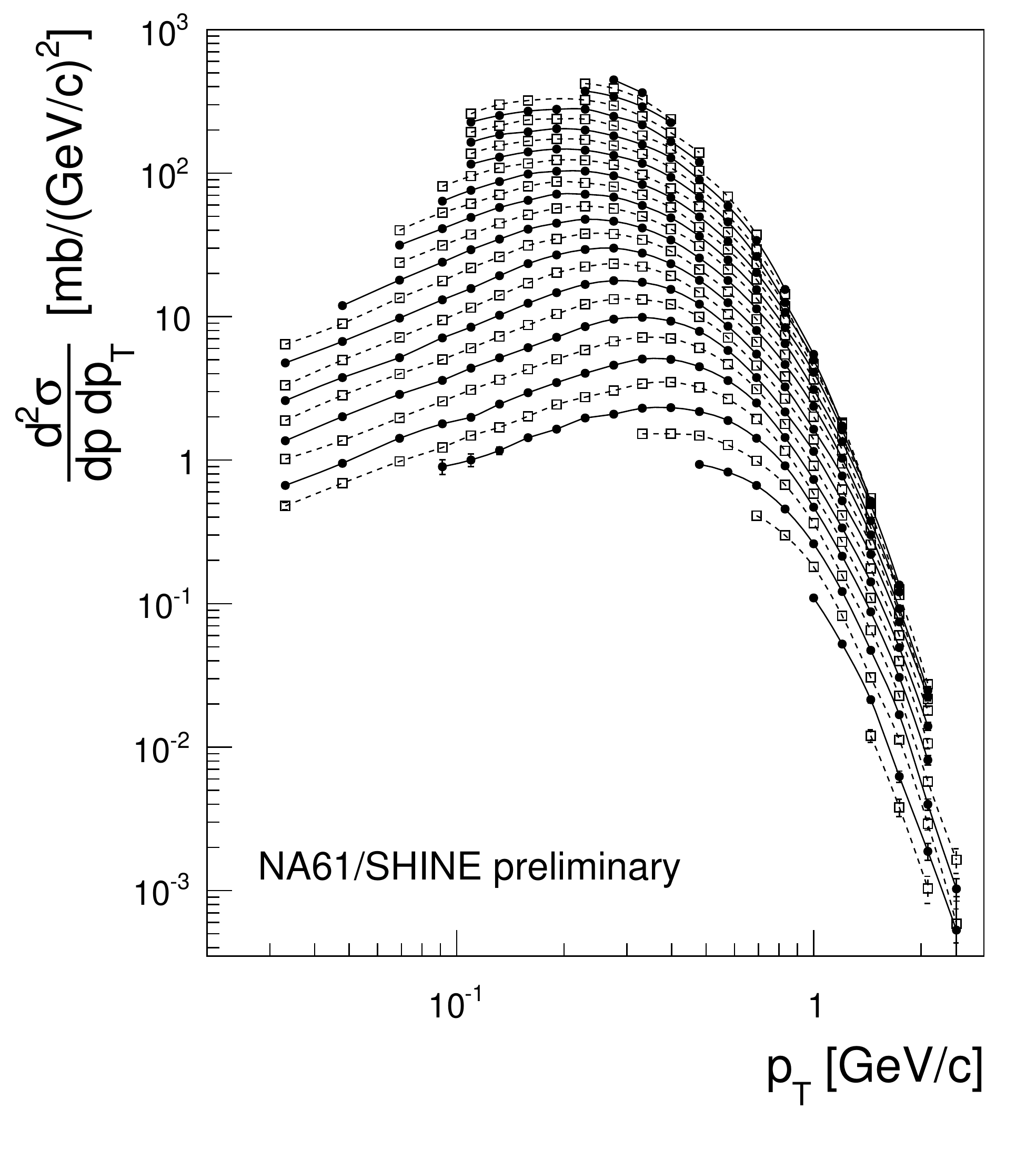}
\label{fig:hMinus158}
}
\subfigure[$h^+$ at 158\,\GeVc]{
\includegraphics[width=0.485\linewidth]{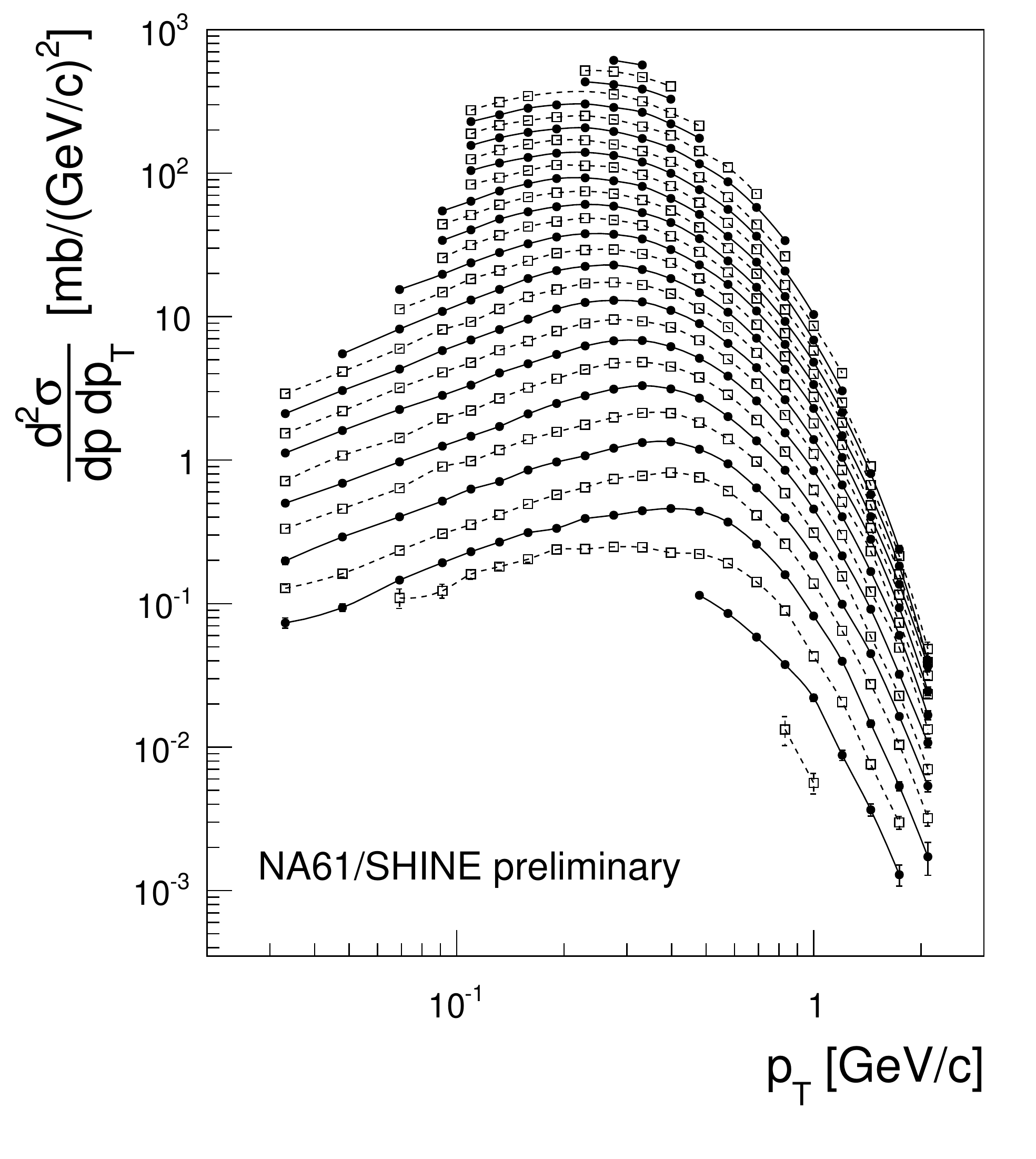}
\label{fig:hPlus158}
}
\subfigure[$h^-$ at 350\,\GeVc]{
\includegraphics[width=0.485\linewidth]{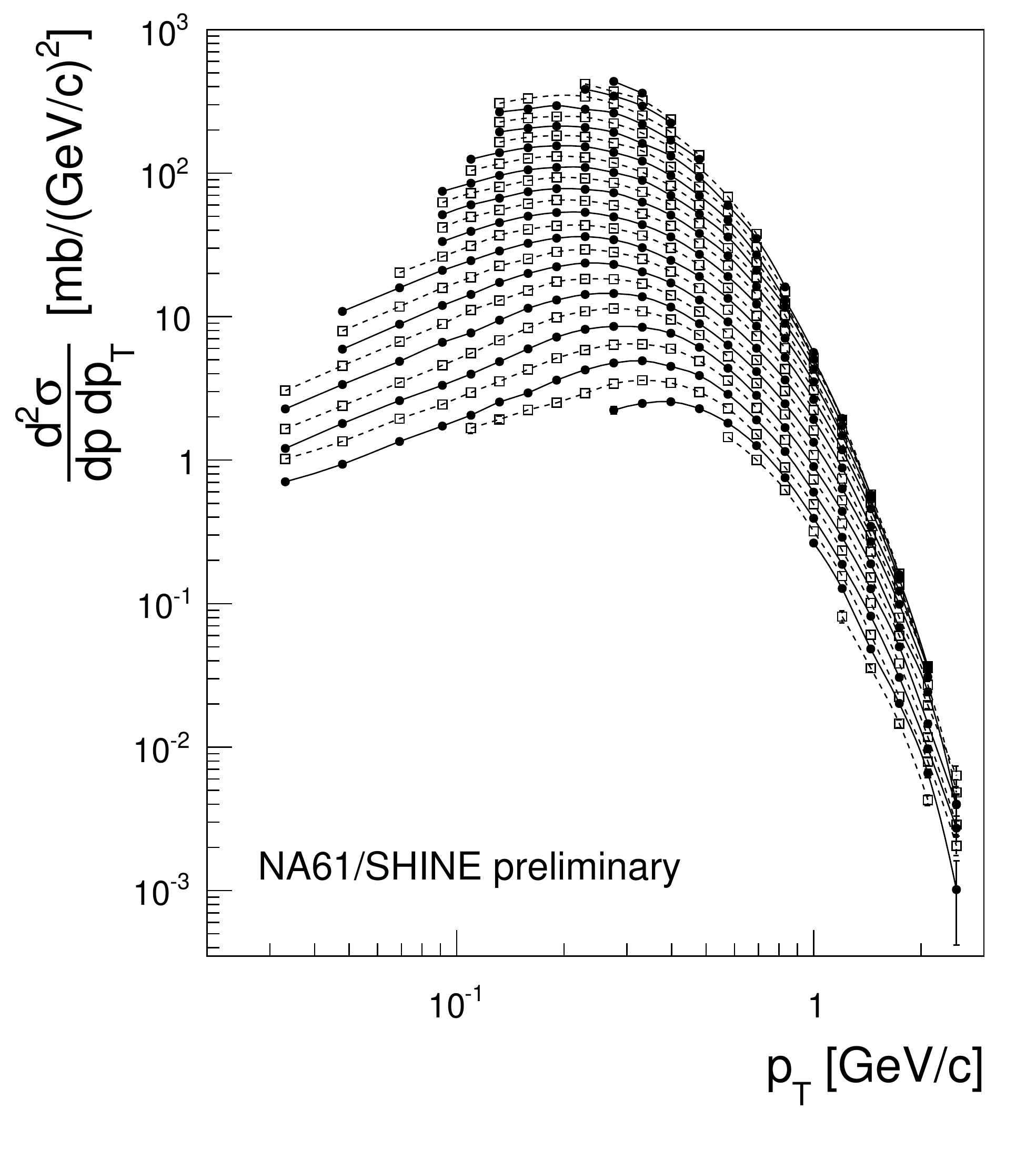}
\label{fig:hMinus350}
}
\subfigure[$h^+$ at 350\,\GeVc]{
\includegraphics[width=0.485\linewidth]{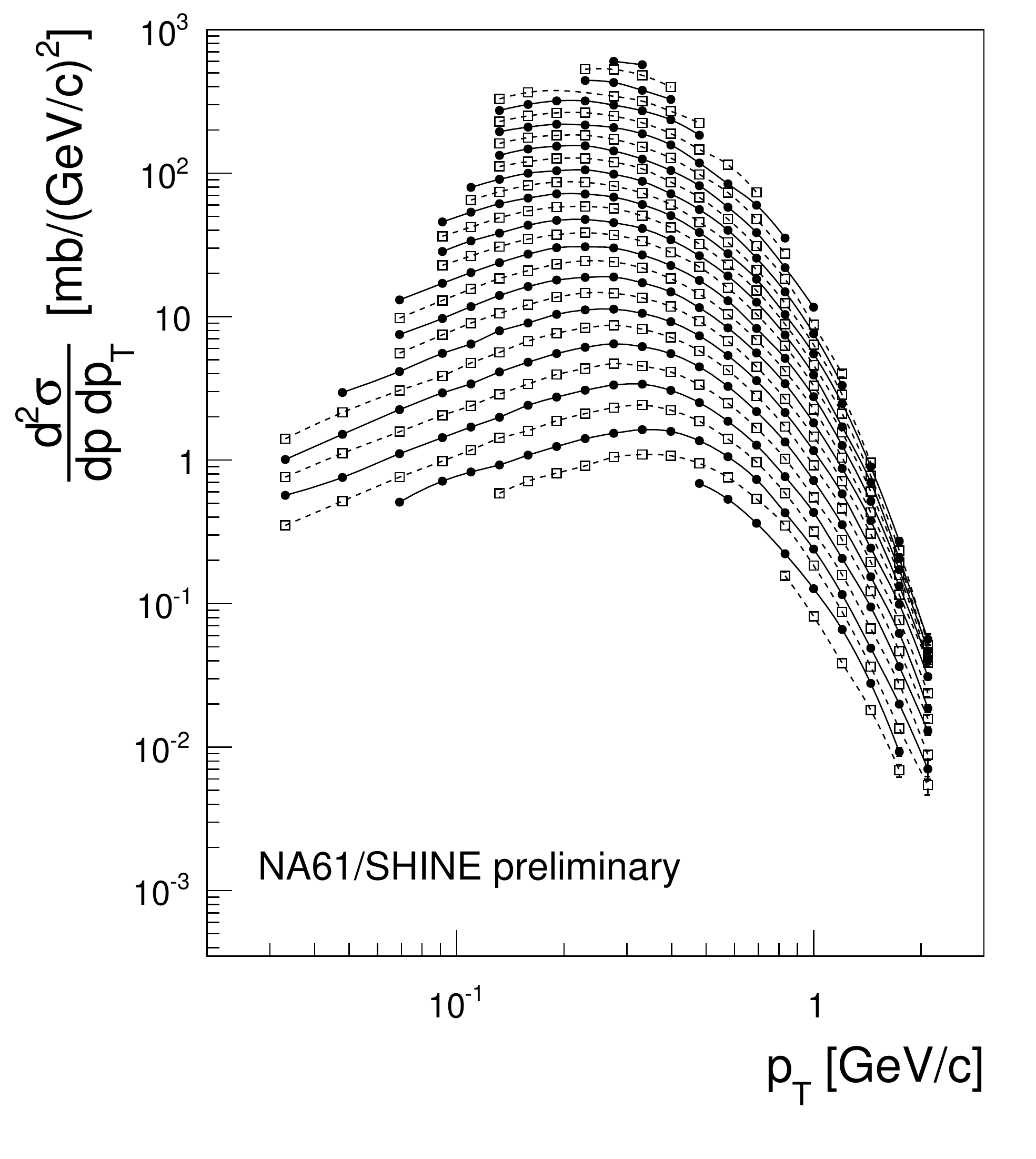}
\label{fig:hPlus350}
}
\caption{Inclusive $p_\text{T}$-spectra of charged hadrons produced in
  $\pi^-$+C interactions at 158 and 350\,\GeVc. In each figure, the
  particle momentum $p$ ranges from 0.6 to 121\,\GeVc in steps of
  $\log(p/(\GeVc)) = 0.08$ from top to bottom.}
\label{fig:CRResults}
\end{figure*}

\begin{figure*}[!t]
\centering
\includegraphics[width=\linewidth]{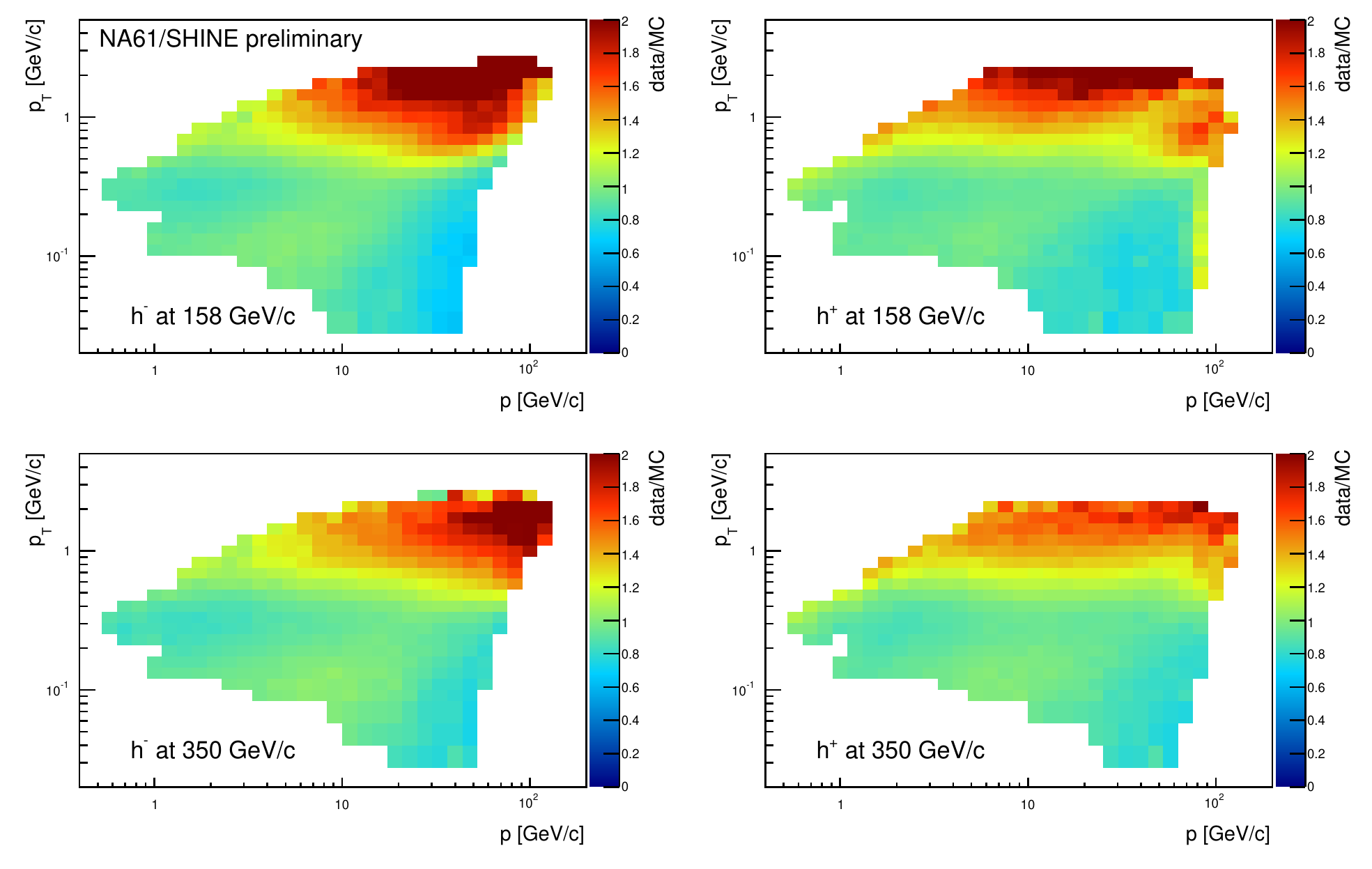}
\caption[CR data MC comparison]{Comparison of measured charged hadron
  production yields in $\pi^-$+C interactions at 158 and 350~\GeVc to
  predictions from \EPOSfull. Colors denote the ratio of data over MC
  and the different panels are for different charges and beam energies
  as indicated by the labels.  Note that the color scale is limited,
  i.e.\ the maximum value to be understood as data/MC$\ge 2$.}
\label{fig:CRDataMC}
\end{figure*}

\begin{figure}[!b]
\centering \includegraphics[clip, viewport= 0 20 550 490,
  width=\linewidth]{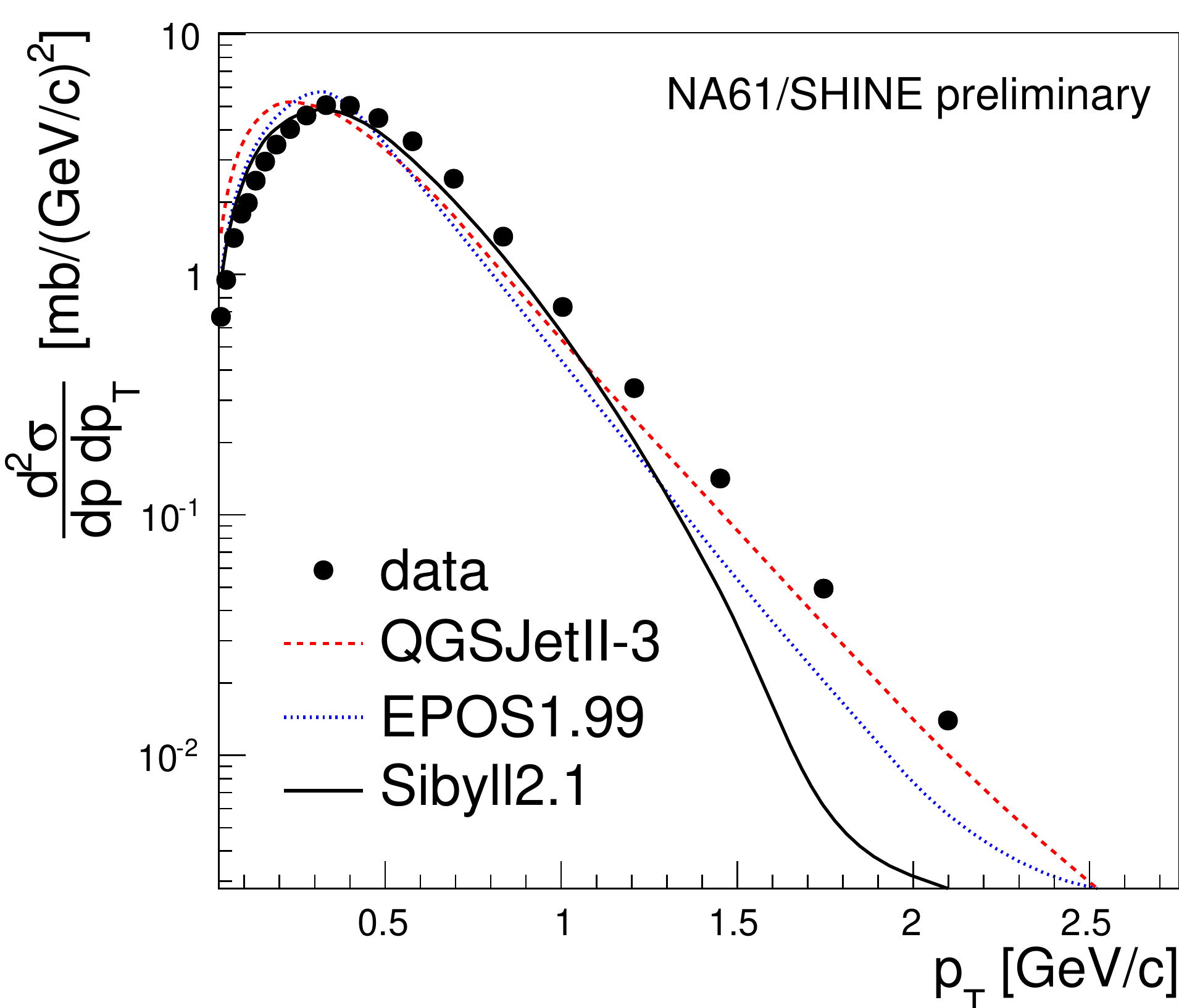}
\caption[CR data MC comparison]{Transverse momentum spectrum of
  negatively charged hadrons produced in $\pi^-$+C interactions at
  158~\GeVc beam momentum at $\langle p\rangle=$10.4~\GeVc.}
\label{fig:CRDataMCExample}
\end{figure}

The production cross section in $\pi^-$+C interactions was determined
in a similar manner as described in Ref.~\cite{NA61pion}, by
correcting the experimental interaction cross section by residual
contributions from elastic and quasi-elastic scattering as well as for
the inelastic contribution to which the NA61/SHINE interaction trigger
is not sensitive. The uncertainties of the measurement is currently
dominated by the model-dependence of this correction.  Preliminary
values are $\sigma_\mathrm{prod} = 172 \pm 2 \,(\mathrm{stat}.) \pm 4
\, (\mathrm{syst.})$ and $178 \pm 2 \,(\mathrm{stat}.) \pm 4 \,
(\mathrm{syst.})$ at 158 and 350~\GeVc respectively~\cite{haugThesis}.
This measurement is compatible with previous
results~\cite{Schiz:1979qf,Dersch:1999zg} and already gives the most
precise value of the production cross section at around 160\,\GeVc.

The momentum spectra of charged hadrons in $\pi^-$+C interactions at
158 and 350\,\GeVc are presented in Fig.~\ref{fig:CRResults}.  These
spectra were obtained within a fiducial phase space in the NA61
detector, for which the detection and selection efficiency for charged
tracks is close to unity, and corrected for feed-down and track loss
using the average correction predicted by the \textsc{Venus} and
\textsc{Epos} event generators after simulation of the detector
response~\cite{ruprechtThesis}.  The trigger bias is corrected for by
studying the track loss in a sub-sample of unbiased beam-trigger
data. Only phase-space regions for which the overall model correction
is below 20\% and for which the total systematic uncertainty is
smaller than 20\% are displayed in Fig.~\ref{fig:CRResults}.  The
uncertainties shown are the total uncertainties including the
statistical uncertainty and systematics from the model correction,
normalization, trigger bias, calibration and track topology.

These preliminary measurements are already useful to judge the quality
of event generators used in air-shower simulations.  An example of the
$p_\mathrm{T}$ distribution of negatively charged hadrons produced in
$\pi^-$+C interactions at a beam momentum of 158~\GeVc is shown in
Fig.~\ref{fig:CRDataMCExample} for particle momenta with $\langle
p\rangle=$10.4~\GeVc and compared to predictions by
\QGSJETIIfull~\cite{qgsjetII}, \SIBYLLfull~\cite{sibyll2.1} and
\EPOSfull~\cite{epos}.  As can be seen, none of these hadronic
interaction models which are used to simulate air showers can
reproduce that data and especially \SIBYLLfull predicts a much too
steep spectrum at high transverse momenta.

The full data set is compared to the predictions of the \EPOSfull
model in Fig.~\ref{fig:CRDataMC} where the ratio of data over MC is
shown.  It can be seen that the underestimation of charged hadron
production at large transverse momenta, which was illustrated in
Fig.~\ref{fig:CRDataMCExample} at one particular momentum, is present
at all momenta (the same holds true for \SIBYLL).  Of all the models
studied \QGSJETIIfull describes our data best with only a small
deficit of tracks with high $p_\mathrm{T}$ at large particle momentum
but slightly too many particles at low traverse momenta.

It is planned to study these shortcomings of the models in more detail
by measuring the spectra of identified hadrons.  Moreover, NA61/SHINE
will be able to validate the measurement of proton and anti-proton
production in $\pi$+C interactions from Ref.~\cite{Barton:1982dg} on
which mostly the enhanced baryon production is based on that has been
proposed in Ref.~\cite{epos} as a possibility to enlarge the number of
muons in air showers. In addition, the NA61/SHINE data set offers the
possibility to constrain the $\rho^0$ production in $\pi$+C which may
be equally important for muons observed in air showers as the baryon
fraction (see e.g.\ Ref.~\cite{Drescher:2007hc}).
\begin{figure*}[t]
\centering \includegraphics[clip, viewport=35 0 490 500,
  width=0.49\linewidth]{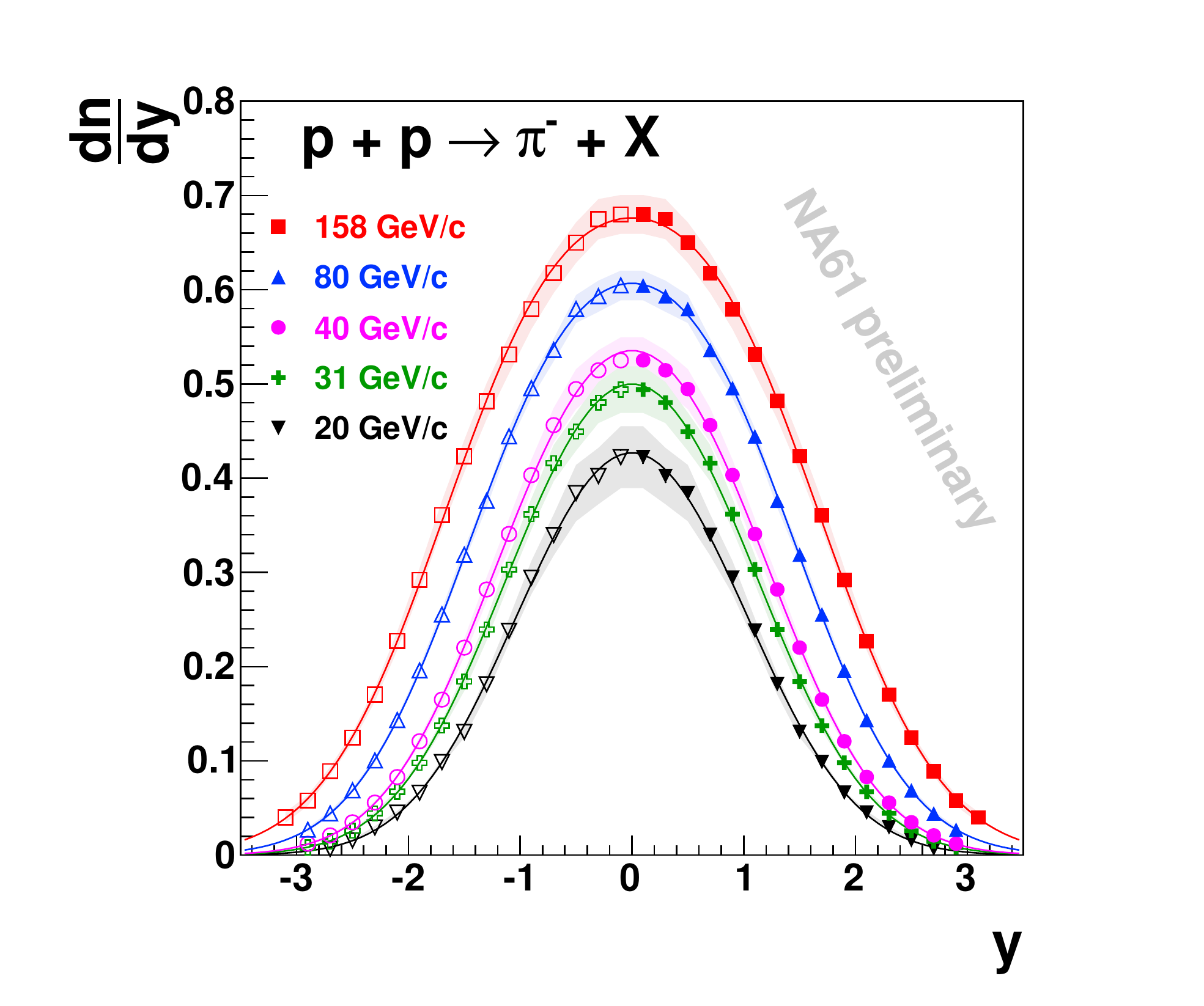}
\includegraphics[clip, viewport=35 0 490 500,
  width=0.49\linewidth]{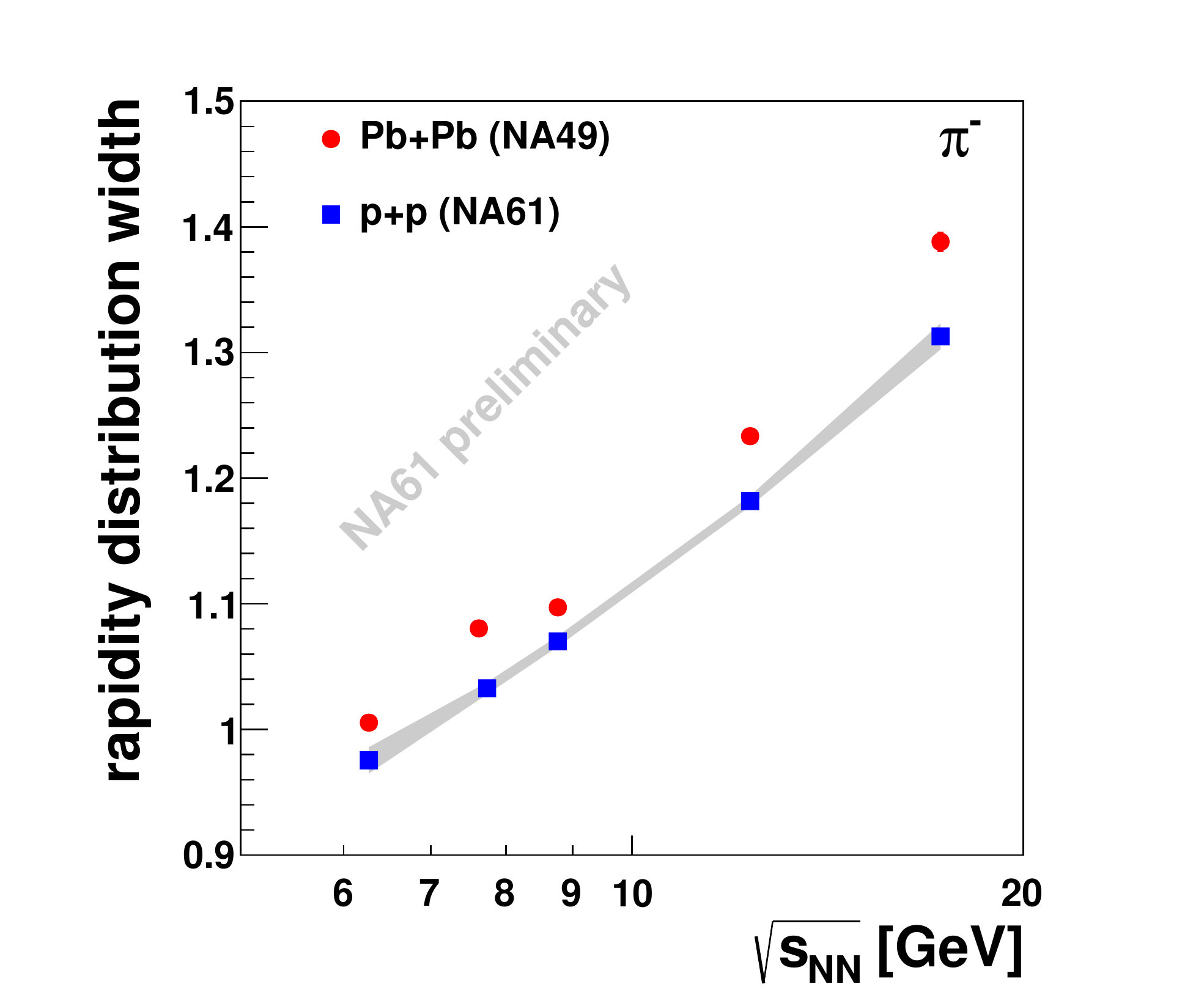}
\caption{\emph{Left:} $m_\text{T}$-integrated $\pi^-$ rapidity spectra
  in $p$+$p$ interactions at 20--158 \GeVc obtained with the $h^-$
  method and fitted with a sum of two symmetrically displaced Gaussian
  functions. Full symbols represent the preliminary measurement and
  open symbols are the same data reflected around zero rapidity.
  \emph{Right:} $\pi^-$ width (RMS) of the rapidity spectrum as a
  function of energy, compared with Pb+Pb data~\cite{Afanasiev:2002mx,
    Alt:2007aa}. }
\label{fig:h-rapidity}
\end{figure*}

\section{The NA61/SHINE Heavy Ion Program}
\begin{figure}[t]
\centering
\includegraphics[width=\linewidth]{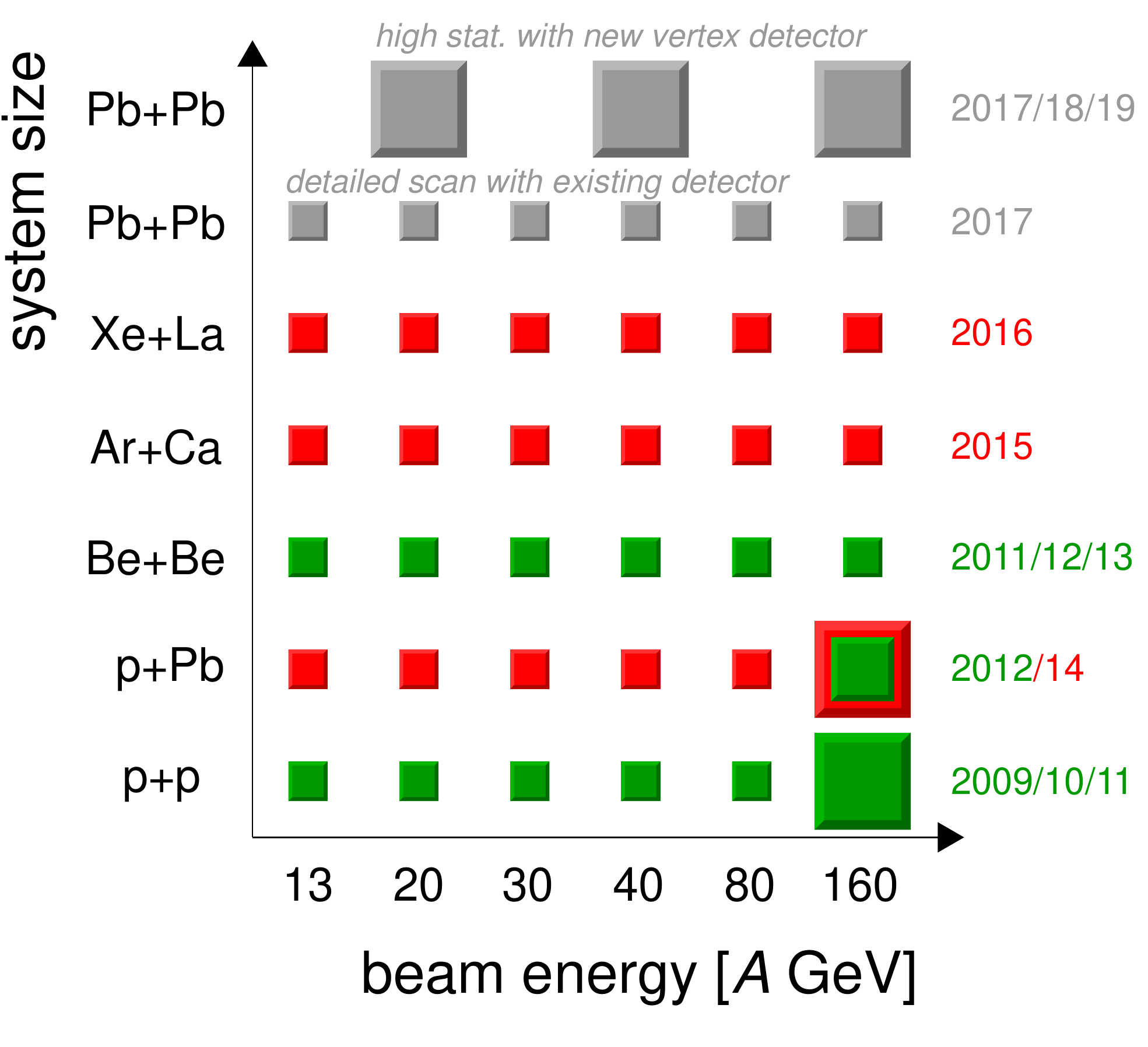}
\caption[runmatrix]{The NA61/SHINE data taking schedule for the ion
  program and its proposed extension for the period 2017--2019. Big
  boxes denote large ($\gtrsim$ 5$\times10^{7}$ events) and small
  boxes medium size ($\sim$1--5$\times10^{6}$ events) data sets. Green
  boxes are recorded data, red boxes denote runs to be taken within
  the approved physics program and gray boxes denote the proposed
  extension.}
\label{fig:na61_plans}
\end{figure}

Within its heavy ion program, NA61/SHINE aims to discover the critical
point of strongly interacting matter as well as to establish the
properties of the onset of deconfinement~(see
Ref.~\cite{Gazdzicki:2010iv} and references therein). The full
experimental program is illustrated in Fig.~\ref{fig:na61_plans} and
consists of a detailed scan of various system sizes and interaction
energies.

As a first step, $p$+$p$ interactions were measured at six energies in
2009-2011 to serve as a reference data set for the subsequent
measurement of light and medium size ion reactions in the range of
$\sqrt{s_{NN}}=$~5--20~\GeV.  In this conference we presented
preliminary spectra of $\pi^-$ in $p$+$p$ collisions at 20, 31, 40,
80, and 158~\GeVc that were obtained using the so-called $h^-$
analysis.  The analysis is based on the fact that the majority of
produced negatively charged particles are pions.  Contribution of
other particles (mostly $K^-$ and \emph{feed-down}) is corrected for
using Monte-Carlo simulations. The corresponding correction is
calculated as arithmetic average of \Venus~\cite{Venus} and
\EPOS~\cite{epos} corrections and the difference between them
contributes to the systematic error.  The detector effects
(acceptance, inefficiency) are corrected for using Monte Carlo as
well.  This approach allows obtaining $\pi^-$ spectra in full measured
phase space in a uniform way.  Non-target interactions,
i.e.~collisions with air and the detector material, are subtracted
using events measured with the empty liquid-hydrogen target.  The
transverse-mass spectra were found to follow an exponential
distribution with $m_T$.  Therefore, they can be extrapolated to full
phase space using an exponential fit to the high $m_T$ tail of the
spectra to obtain the $m_T$-integrated rapidity spectra presented in
Fig.~\ref{fig:h-rapidity} (left).  The spectra are well described by a
sum of two symmetrically displaced Gaussian functions.  The widths of
the rapidity spectra are presented in Fig.~\ref{fig:h-rapidity}
(right) and compared to Pb+Pb data.\\

In addition to these results, preliminary inclusive spectra of
identified pions, protons and kaons are also available from NA61/SHINE
that were obtained by using the energy deposit in the TPCs for
particle identification. With the help of the 'identity'-method
described in Ref.~\cite{Gazdzicki:2011xz}, estimates of multiplicity
fluctuations in $p$+$p$ interactions could be given.  These
measurements will form the reference data set for the full ion
program. As can be seen in Fig.~\ref{fig:na61_plans}, NA61/SHINE
already finished the energy scan with light Be ions that were obtained
by fragmenting primary Pb ions from the
SPS~\cite{Strobele:2012zz}. Data taking will resume with $p$+Pb
interactions in 2014 and the remaining two system sizes, Ar+Ca and
Xe+La, will be measured in 2015 and 2016. Within a possible extension
of the approved physics program, it is foreseen to provide more Pb+Pb
data at six energies and to study $D$-meson production in heavy ion
collisions at SPS energies with high statistics runs and a new vertex
detector~\cite{add6}.

\section{Conclusions}
In this article, we summarized results from the multi-purpose
experiment NA61/SHINE at the CERN SPS.  In the light of the main focus
of this conference, Very High Energy Cosmic Ray Interactions, it is
worth noting that although the different measurements are motivated by
different physics goals, all the hadron-nucleus and proton-proton data
collected by NA61/SHINE are valuable for the tuning of hadronic
interaction models for the understanding of air showers.  The
published NA61/SHINE data on $p$+C interactions at
31~\GeVc~\cite{NA61pion} have already been used to
fix~\cite{Uzhinsky:2011ir} the \UrqmdLong model~\cite{Urqmd} and to
further strengthen~\cite{MUatICHEP} the case against the use of
\Gheisha~\cite{Gheisha}. Both of these models are available in
\Corsika~\cite{Corsika} to simulate low energy interactions in air
shower.  Moreover, recent comparisons~\cite{SzymonCPOD} of NA61/SHINE
data to \EPOS suggest that high-energy interaction models need further
tuning even for 'well-known' reactions like $p$+$p$.

Together with the special cosmic ray runs for pion-carbon
interactions, the rich data set of NA61/SHINE will thus be very useful
to reduce uncertainties in air shower calculations and
increase the knowledge on interactions in the late stages of the
shower development below a \TeV.



\begin{thebibliography}{}
\bibitem{na61} N.~Antoniou {\em et al.} [NA61/SHINE Collaboration],
  CERN-SPSC-2007-004, (2007); CERN-SPSC-2007-019, (2007); N.~Abgrall
  {\em et al.} [NA61/SHINE Collaboration], CERN-SPSC-2008-018, (2008).

\bibitem{Abe:2011ks} K.~Abe {\it et al.}  [T2K Collaboration],
  Nucl.\ Instrum.\ Meth.\ A {\bf 659} (2011) 106.

\bibitem{Gazdzicki:2010iv} M.~Gazdzicki, M.~Gorenstein, P.~Seyboth,
  Acta Phys.\ Polon.\ {\bf B42 } (2011) 307.

\bibitem{Afanasev:1999iu} S.~Afanasev {\it et al.}  [NA49
  Collaboration], Nucl.\ Instrum.\ Meth.\ A {\bf 430} (1999) 210.

\bibitem{Ivashkin:2012fd} A.~Ivashkin {\em et al.}, arXiv:1205.4864.

\bibitem{NA61pion} N.~Abgrall {\it et al.}  [NA61/SHINE
  Collaboration], Phys.\ Rev.\ C {\bf 84} (2011) 034604.

\bibitem{Abgrall:2011ts} N.~Abgrall {\it et al.}  [NA61/SHINE
  Collaboration], Phys.\ Rev.\ C {\bf 85} (2012) 035210.

\bibitem{Abe:2011sj} K.~Abe {\it et al.}  [T2K Collaboration],
  Phys.\ Rev.\ Lett.\ {\bf 107} (2011) 041801; Phys.\ Rev.\ D {\bf 87}
  (2013) 012001.

\bibitem{Venus}
K.~Werner, Nucl. Phys. A525, 501c (1991); Phys. Rep. 232, 87 (1993).

\bibitem{Fluka} A.~Fasso {\em et al.}, CERN-2005-10 (2005);
  G. Battistoni {\em et al.}, AIP Conf.\ Proc.\ {\bf 896} (2007) 31.

\bibitem{Uzhinsky:2011ir} V.~Uzhinsky, arXiv:1107.0374 [hep-ph].

\bibitem{Urqmd} S.M.~Bass {\em et al.}, Prog. Part. Nucl. Phys.  {\bf
  41} 225 (1998); M.~Bleicher {\em et al.}, J.  Phys. G:
  Nucl. Part. Phys. {\bf 25}, 1859 (1999).

\bibitem{Abgrall:2012pp} N.~Abgrall {\it et al.}  [NA61/SHINE
  Collaboration], Nucl.\ Instrum.\ Meth.\ A {\bf 701} (2013) 99.

\bibitem{Auger} J.~Abraham {\em et al.} [Pierre Auger Collaboration],
  Nucl.\ Instrum.\ Meth.\ A {\bf 523} (2004) 50.

\bibitem{KASCADE} T.~Antoni {\em et al.} [KASCADE Collaboration],
  Nucl.\ Instrum.\ Meth.\ A {\bf 513} (2003) 490.

\bibitem{IceTop} R.~Abbasi {\it et al.}  [IceCube Collaboration],
  Nucl.\ Instrum.\ Meth.\ A {\bf 700} (2013) 188.

\bibitem{Heck:2003br} D.~Heck {\em et al.}, Proc.\ 28th ICRC, (2003)
  279.

\bibitem{Drescher:2003gh} H.-J.~Drescher {\em et al.},
  Astropart.\ Phys.\ {\bf 21 } (2004) 87.

\bibitem{Meurer:2005dt} C.~Meurer {\em et al.},
  Czech.\ J.\ Phys.\ {\bf 56 } (2006) A211.

\bibitem{Maris:2009uc} I.~C.~Mari\c{s} {\em et al.},
  Nucl.\ Phys.\ Proc.\ Suppl.\ {\bf 196 } (2009) 86.

\bibitem{Maris:2009x1} I.~C.~Mari\c{s} {\em et al.} [NA61/SHINE
  Collaboration], Proc.\ 31st ICRC, (2009).

\bibitem{haugThesis} M.\ Haug, {\it Messung des Wirkungsquerschnitts
  von Pion-Kohlenstoff-Wechselwirkungen mit Hilfe des NA61 Detektors},
  Diploma Thesis, KIT, 2012.

\bibitem{ruprechtThesis} M.\ Ruprecht, {\it Measurement of the
  Spectrum of Charged Hadrons in $\pi^-$+C Interactions with the NA61
  Experiment}, Diploma Thesis, KIT, 2012.

\bibitem{epos} T.\ Pierog and K.\ Werner, Phys.\ Rev.\ Lett.\ {\bf
  101} (2008), 171101.

\bibitem{sibyll2.1} E.J.~Ahn \emph{et al.}, Phys.\ Rev.\ {\bf D80 }
  (2009) 094003.

\bibitem{qgsjetII} S.S. Ostapchenko, Nucl.\ Phys.\ Proc.\ Suppl.\ {\bf
  151} (2006), 143;

\bibitem{Schiz:1979qf} A.~Schiz {\it et al.}, Phys.\ Rev.\ D {\bf 21}
  (1980) 3010.

\bibitem{Dersch:1999zg} U.~Dersch {\it et al.}\ [SELEX Collaboration],
  Nucl.\ Phys.\ B {\bf 579} (2000) 277.

\bibitem{Barton:1982dg} D.~S.~Barton {\it et al.}, Phys.\ Rev.\ D {\bf
  27} (1983) 2580.

\bibitem{Drescher:2007hc} H.~-J.~Drescher, Phys.\ Rev.\ D {\bf 77}
  (2008) 056003.

\bibitem{Afanasiev:2002mx} S.~V.~Afanasiev {\it et al.}  [NA49
  Collaboration], Phys.\ Rev.\ C {\bf 66} (2002) 054902.

\bibitem{Alt:2007aa} C.~Alt {\it et al.}  [NA49 Collaboration],
  Phys.\ Rev.\ C {\bf 77} (2008) 024903.

\bibitem{Gazdzicki:2011xz} M.~Gazdzicki, K.~Grebieszkow, M.~Mackowiak
  and S.~Mrowczynski, Phys.\ Rev.\ C {\bf 83} (2011) 054907

\bibitem{Strobele:2012zz} H.~Str\"obele and I.~Efthymiopoulos,
  CERN Cour.\  {\bf 52N4} (2012) 33.

\bibitem{add6} N.~Abgrall {\it et al.}\ [NA61/SHINE Collaboration],
  {\it NA61/SHINE plans beyond the approved program},
  CERN-SPSC-2012-022; SPSC-P-330-ADD-6.

\bibitem{Gheisha} H.~Fesefeldt, Aachen Report No.~PITHA-85-02 (1985)

\bibitem{MUatICHEP} M.~Unger for the NA61/SHINE~Collaboration,
  Proc.\ 35th ICHEP (2010), arXiv:1012.2604.

\bibitem{Corsika} D.~Heck {\em et al.}, FZK Report No.~FZKA-6019,
  1998.

\bibitem{SzymonCPOD} S.~Pulawski for the NA61/SHINE~Collaboration,
  talk at the 8th CPOD (2013).

\end{thebibliography}
\end{document}